%% file: CANDAR2024_Main.tex
\documentclass[10pt, conference, compsocconf]{IEEEtran}
\usepackage{amsmath,amssymb,amsfonts}
\usepackage{graphicx}
\usepackage{color}
\usepackage[caption=false,font=footnotesize]{subfig}
\usepackage{booktabs}
\usepackage{multirow}
\usepackage{setspace}
\usepackage{algpseudocode} %Adding by Harie
\usepackage{ltablex}
\usepackage{geometry}
\usepackage{float}
\usepackage{afterpage}
\usepackage[linesnumbered,ruled,vlined]{algorithm2e}
\usepackage{subcaption}
\usepackage{hyperref}
\SetKwComment{Comment}{$\triangleright$\ }{}
\SetKwInput{KwInit}{Initialization}
\makeatletter
\newcommand{\linebreakand}{%
  \end{@IEEEauthorhalign}
  \hfill\mbox{}\par
  \mbox{}\hfill\begin{@IEEEauthorhalign}
}
\makeatother

\begin{document}
\title{CodoMo: \\Python Model Checking to Integrate Agile Verification Process of Computer Vision Systems}


\author{
\IEEEauthorblockN{Yojiro Harie}
\IEEEauthorblockA{
    Department of Information Engineering\\
    Kanazawa Gakuin University\\
    Kanazawa, Ishikawa, Japan\\
    Email: \href{mailto:harie@kanazawa-gu.ac.jp}{harie@kanazawa-gu.ac.jp}}
\and
\IEEEauthorblockN{Yuto Ogata}
\IEEEauthorblockA{
    Department of Information Engineering\\
    Kanazawa Gakuin University\\
    Kanazawa, Ishikawa, Japan\\
    Email: y-ogata@kanazawa-gu.ac.jp}
\linebreakand
\IEEEauthorblockN{Bishnu Prasad Gautam}
\IEEEauthorblockA{
    Department of Information Engineering\\
    Kanazawa Gakuin University\\
    Kanazawa, Ishikawa, Japan\\
    Email: gautam@kanazawa-gu.ac.jp}
\and
\IEEEauthorblockN{Katsumi Wasaki}
\IEEEauthorblockA{
    Faculty of Engineering\\
    Shinshu University\\
    Nagano, Nagano, Japan\\
    Email: wasaki@cs.shinshu-u.ac.jp}
}

\maketitle

\input{Ch0_Abstract_KeyWords_Harie}
\input{Ch1_Intro_Harie} % task of Harie 
\input{Ch2_Background} % task of gautam
\input{Ch3_Proposed_Solution_Harie} % task of Harie
\input{Ch4_Impl_Details_HO} % task of Harie and Ogata
\input{Ch5_Result_Discussion_HOG} % task of Harie, Ogata and Gautam
\input{Ch6_Conclusion_HOG} % task of Harie, Ogata and Gautam
\input{Ch7_Acknowledgement} % task of gautam

%\bibliographystyle{IEEEtran}
%\bibliography{ref}

% Consolidated Bibliography
%\printbibliography
%\usepackage[style=ieee]{biblatex}

%\end{document}

%\addbibresource{ref.bib}
%\addbibresource{Lit-Rev.bib}

% \nocite{*} % This will print all entries in bibliography file
%\printbibliography

\bibliographystyle{IEEEtran} % BibTeXスタイル指定
\bibliography{ref} % 参照ファイル名（拡張子不要）

\end{document}

%% file: Ch0_Abstract_KeyWords_Harie.tex
\begin{abstract}
Model checking is a fundamental technique for verifying finite state concurrent systems.
Traditionally, model designs were initially created to facilitate the application of model checking.
This process, representative of Model Driven Development (MDD), involves generating an equivalent code from a given model which is verified before implementation begins.
However, this approach is considerably slower compared to agile development methods and lacks flexibility in terms of expandability and refactoring.
We have proposed ``CodoMo: Python Code to Model Generator for pyModelChecking.''
This tool automates the transformation of a Python code by an AST static analyzer and a concolic testing tool into intermediate models suitable for verification with pyModelChecking, bridging the gap between traditional model checking and agile methodologies.
Additionally, we have implemented a multiprocess approach that integrates the execution of PyExZ3 with the generation of Kripke structures achieving greater work efficiency.
By employing CodoMo, we successfully verified a Tello Drone programming with gesture-based image processing interfaces, showcasing the tool’s powerful capability to enhance verification processes while maintaining the agility required for today’s fast-paced development cycles.

\end{abstract}

\begin{IEEEkeywords}
Model-Driven Reverse Engineering; concolic testing; Agile Integration; Model Checking; CodoMo 
\end{IEEEkeywords}

\IEEEpeerreviewmaketitle

%% file: Ch1_Intro_Harie.tex
\section{Introduction}
Formal specifications and model checking are traditional mathematical methods for proving whether critical complex systems have no error.  For example, Amazon Web Services (AWS) utilizes two model checking tools, PlusCal for designing robust distributed systems to build fault-tolerant low-level network algorithms, and TLA+ for designing database replication systems to manage high traffic loads \cite{amazon2015}.

Model checking enables exhaustive error detection that code reviews and static analysis may miss.
However, applying  model checking methods to real-world systems presents challenges, such as the state explosion problem and  potential misalignment with agile software development practices.
Model-driven development (MDD), which incorporates formal methods from the development stage and verifies them according to the development granularity, has been proposed.
MDD is a software engineering approach that applies models and model technologies to raise the level of abstraction at which developers create and evolve software cite{hailpern2006model}.
MDD is composed of various phases, including use case analysis, activity definition, requirement validation, formal verification, and model transformation.

Since MDD typically involves defining models in the initial stages of system development and relies on model transformations based on these predefined models, it becomes challenging to adapt to agile development, where changing requirements often lead to evolving specifications.
As a result, updating models flexibly as development progresses becomes difficult.
Consequently, integrating agile development with Model-Driven Engineering (MDE) has been identified as a significant challenge \cite{bucchiarone2020grand}.
Additionally, legacy large-scale systems often lack architectural and design documentation, and their specifications and data flow may not be self-explanatory.
Whereas, Model-driven reverse engineering (MDRE) draws inspiration from traditional reverse engineering (RE) and was developed to reconstruct models from systems, generating model-based views of legacy systems \cite{sabir2019model}.
However, while MDRE focuses on extracting models from existing systems, methodologies and tools that effectively incorporate formal verification into the ever-changing environment of agile development are still lacking.
\begin{figure*}[t]
\begin{center}
  \includegraphics[width=\linewidth]{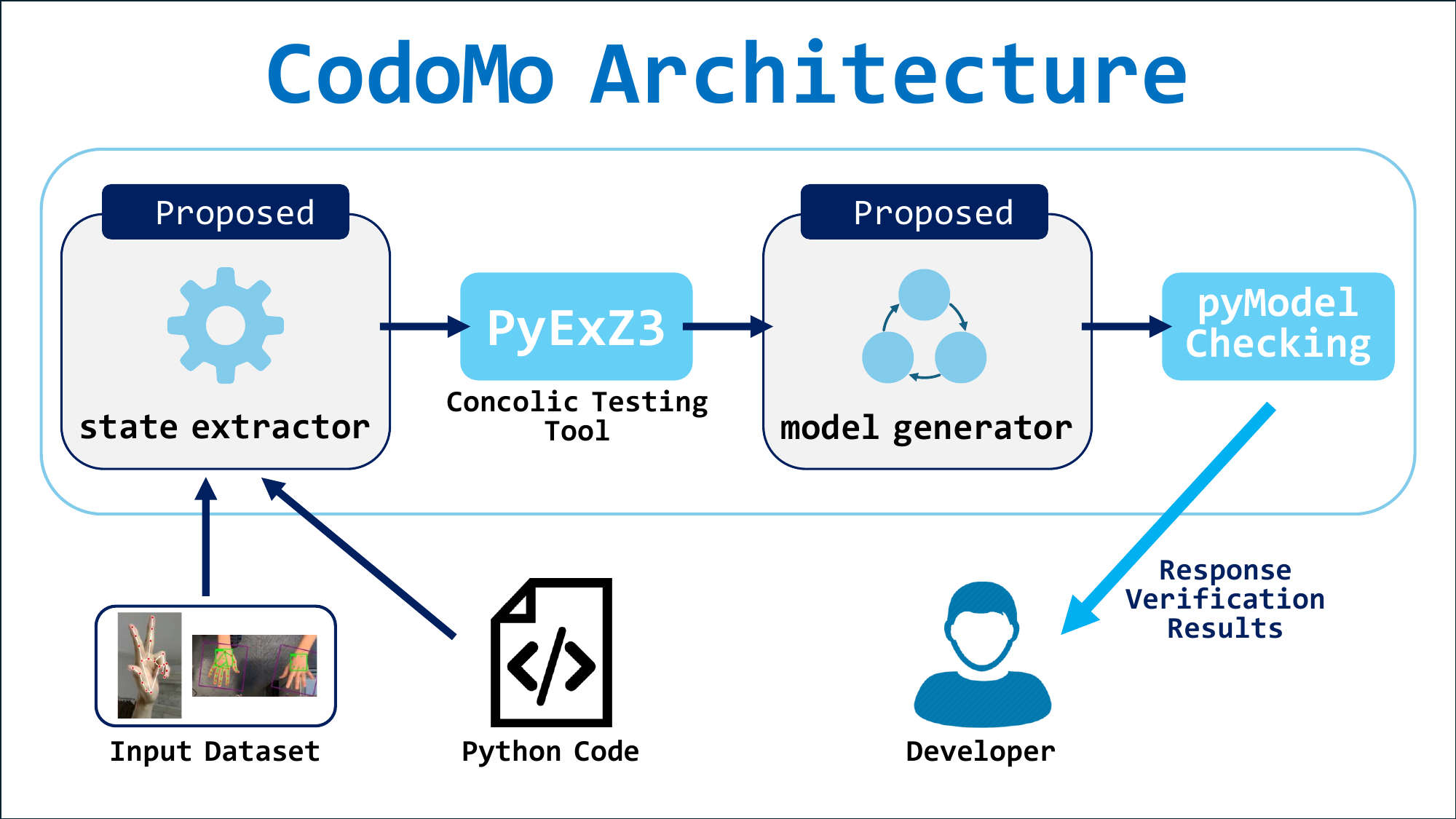}
  \caption{CodoMo system architecture}
  \label{figure1}
  \end{center}
\end{figure*}

This paper presents ``CodoMo'' tool designed to integrate model checking and the dynamism of agile development for structured systems.
CodoMo aims to integrate the precision of model checking while maintaining the flexibility required by agile methodologies.
We propose the method involves seamless integration of the PyExZ3 concolic testing and pyModelChecking tools (see Fig.\ref{figure1}.)
CodoMo automates the transformation of Python code into intermediate models suitable for verification with pyModelChecking.
This integration can generate and verify models that are equivalent to Python implementations.

The process begins by inputting both a target Python code and the corresponding input image set into CodoMo tool.
CodoMo then semi-automatically converts the Python code for use with PyExZ3, where it conducts concolic testing to explore output transition paths.
From these paths, state transition information is extracted and subsequently transformed into Kripke structures.
We have also introduced a data-driven approach, using a dataset of actual captured images and videos to generate intermediate models that closely reflect real-world conditions.

By leveraging an Abstract Syntax Tree (AST) static analyzer and a concolic testing tool, CodoMo bridges the gap between model-driven approaches and agile practices, enabling more efficient and flexible verification processes.
This approach allows developers to maintain the agility of their workflow while ensuring the software-correctness through formal verification.

Moreover, CodoMo incorporates a multi-process approach that integrates the execution of PyExZ3 with the generation of Kripke structures, which significantly enhances work efficiency.
As an application example of CodoMo, we have successfully verified a Tello Drone programming \cite{ryze2024}, particularly focusing on gesture-based image processing interfaces
This demonstrates its ability to integrate verification methods with agile methodologies, ensuring robustness and reliability in real-world applications, even for complex systems.

In the literature, we refer to examples of concolic testing for drone control systems using scenario models have already been presented \cite{Katz2021,Katz2019}.

This article is organized as follows.
Section \ref{sec:2} provides explanations of terms related to CodoMo tool.
In Section \ref{sec:3}, we introduce the main architecture of CodoMo tool.
The details of our implementation are presented in Section \ref{sec:4}.
In Section \ref{sec:5}, we describe the experiments conducted using model checking on an interfacing system that analyzes hand gestures with computer vision technology to control a drone.
Finally, in Section \ref{sec:6}, we discuss future prospects and challenges, and conclude the paper.

%% file: Ch2_Background.tex
\section{Background}
\label{sec:2}
\subsection{Concolic Testing}
The symbolic execution is a verification technique that explores all possible paths through the program by collecting the constraints at branching points \cite{anand2008demand}.
Dynamic symbolic Execution (DSE) is a form of path-based symbolic execution.
As DSE combines both concrete and symbolic reasoning, it also has been called ``concolic'' testing \cite{ball2015deconstructing}.
Concolic testing executes the program on some given or random inputs and collects the constraints of the branches that were followed during execution.
The constraint solver finds a value satisfying conditions of a path in branches. Then, the other paths are obtained by the negation of that conditions.
This allows the solver to infer a new input for the next execution, forcing the program to explore an unexplored path.
Concolic testing is widely used in unit testing because it can quickly detect bugs and improve test coverage in a relatively smaller search space.

\subsection{PyExZ3}
Prominent symbolic execution tools include Java PathFinder (JPF) \cite{havelund2000model}, which can analyze Java programs, and Z3 \cite{de2007efficient}, which can analyze Python code.
JPF developed by NASA's research labs, uses symbolic execution to detect deadlocks and race conditions and it has been utilized in the verification of control systems for Mars rovers.
Z3 is a solver based on Satisfiability Modulo Theories (SMT), which generalizes boolean satisfiability (SAT) by incorporating additional reasoning capabilities such as equality reasoning, arithmetic, fixed-size bit-vectors, arrays, quantifiers \cite{de2008z3}.

PyExZ3 \cite{ball2015deconstructing} is a concolic testing tool, written in Python, that uses the Z3 solver to determine concrete values of variables for each branch condition in Python programs.
The PyExZ3 tool can either perform exhaustive testing of an entire Python code or focus on a specific function by designating it.
When a function is designated, the arguments of the function are targeted for the tracking in the concolic testing.
The tool then solves the assignments related to the branch conditions associated with those arguments and systematically explores all possible execution paths within the function.

\subsection{Model Checking}
Model checking is an automatic verification technique that evaluates a finite-state system against temporal logic formulas, reducing the problem to a graph exploration process \cite{clarke1997model}.
In automata-theoretic model checking, the design under verification is composed of a B\"{u}chi automaton that accepts traces violating the specification.
In model checking, temporal logics such as Computational Tree Logic (CTL), Linear Temporal Logic (LTL), and their combination, CTL* (CTL star), are used.
The system is represented by a Kripke structure, which is then composed with the specification described in temporal logic.

The formal syntax of CTL* is given below. 
Let $AP$ be a set of atomic propositions.
Any CTL* formula is either a state formula or a path formula.
A CTL* state formula is either:
\begin{itemize}
\setlength{\itemsep}{1mm}
\item $p \in AP$;
\item if $f$ and $g$ are state formulas, then $\neg f,$ $f\lor g,$ $f \land g$, or $f \rightarrow g$ is a state formula;
\item if $f$ is a path formula, then $\mathbf{E}f$ and $\mathbf{A}f$.
\end{itemize}
A CTL* path formula is either:
\begin{itemize}
\setlength{\itemsep}{1mm}
\item every state formula;
\item if $f,$ $g$ are path formulas, then  $\neg f,$ $f\lor g,$ $f \land g$, $\mathbf{X}$$f$, $\mathbf{G}$$f$, $\mathbf{F}$$f$, $f\mathbf{U}g$ and $f\mathbf{R}g$ are path formulas.
\end{itemize}

The semantics of CTL* formula is given with respect to a Kripke structure.
The semantics is described in \cite{clarke1997model}.
Let $AP$ be the set of atomic propositions.
Formally, a Kripke structure is ordered tuple, $M :=\langle S, R, P, S_0 \rangle $, where
\begin{itemize}
\setlength{\itemsep}{1mm}
\item $S$ is a finite set of states,
\item $R$ is the binary relation on $S$ which gives the possible transitions between states,
\item $P$ is assignment of atomic propositions of states, that is, $P:S\rightarrow2^{AP}$,
\item $S_0\subseteq S,$ is the set of initial states.
\end{itemize}

\subsection{pyModelChecking}
pyModelChecking \cite{casagrande_pyModelChecking} is an open-source model checker implemented in Python, designed to verify temporal logic properties of finite-state systems.
pyModelChecking includes a parser capable of analyzing formulas expressed as strings in CTL, LTL, and CTL*.
pyModelChecking allows users to define Kripke structures directly in Python code, facilitating the modeling and verification of systems without the need for external tools. 

\subsection{Reverse-Modeling Tool Comparison}
JPF \cite{havelund2000model} can perform reverse modeling from Java program code to verify the overall execution path using input data of integer and string types.
However, JPF is inherently limited in its ability to handle image data as input and lacks the capability to generate models based on such non-standard data types.

ESBMC (Efficient SMT-Based Context-Bounded Model Checker) can verify C++ code using symbolic execution, effectively handling modern C++ features such as polymorphism and template types \cite{ESBMC2023}.
Additionally, ESBMC's capabilities in model checking C++ programs, including its support for CUDA, could theoretically be extended to verify CUDA-based computer vision algorithms. 
In fact, ESBMC has been applied for verifying ANNs \cite{ESBMC_AI}, including the detection of adversarial examples and the verification of coverage methods in multi-layer perceptrons (MLP).
However, these verifications were conducted on 5x5 pixel character images, which means that ESBMC is not yet fully equipped to handle complex computer vision systems.

PyModel \cite{jacky2011pymodel} is a framework that generates finite state machines (FSMs) from Python code by leveraging model-based testing concepts.
PyModel combines these models to generate test cases, which are then executed to verify whether the system behaves according to the model.
This process helps ensure consistency between the specifications and the implementation, confirming that the system operates as expected.
PyModel does not have the capability to verify the behavior of Python program code; rather, it is a tool designed to generate and verify FSMs based on Python code written in PyModel's specific notation.

The comparison of the above-mentioned tools highlights the following features of the proposed CodoMo tool.

\textbf{Advantages of CodoMo}
\begin{itemize}
    \item CodoMo can verify system behaviors using input image data and is commonly adopted in computer vision systems developed with Python.
    \item By adopting a parallel concolic testing architecture, CodoMo can be applied to on-the-fly model checking.
\end{itemize}

\textbf{Limitations of CodoMo}
\begin{itemize}
    \item CodoMo does not address the state space explosion problem as opposed to one in Bounded Model Checking.
    \item CodoMo requires partial human-written code conversion for MDRE and model checking.

\end{itemize}

%% file: Ch3_Proposed_Solution_Harie.tex
\section{CodoMo: Python Code to Model Generator for PyModelChecker}
\label{sec:3}

%We now provide details on the state extractor, PyExZ3, model generator, and PyModelChecking (see Fig.\ref{figure1}.)

This section explains the main components of CodoMo tool, which semi-automates the process of extracting models from code and performs model checking.
Specifically, we will discuss the State Extractor part and Model Generator components.
Detailed implementation of these parts is provided in Section~\ref{sec:4}.
\subsection{State Extractor}

%%addded by Harie
We focused on extracting state transitions from the initial variables within class constructors.
Specifically, for this drone programming, class attributes are critical observational points used to control drones through hand gestures based on its state.
We achieved limited modeling that enables the generation of exhaustive trace graphs of class variables, from initialization to the final state, using concolic execution.

While PyExZ3 excels at concolic testing on functions, it lacks the capability to directly verify class libraries.
To address this limitation, we developed a method that transforms Python class structures into a format compatible with PyExZ3.
This is done by converting a given Python code into the form of AST and performing a reverse transformation from the AST back into another code, enabling PyExZ3 to process class libraries effectively.
Also, PyExZ3 tracks the arguments of the designated functions, whereas we may have to extract states from other parameters in the functions.
Thus, state extractor can replace the arguments with designated attributes or parameters in the functions and rearrange them based on the AST without changing its original context.

Furthermore, we inserted program statements into the target functions to load images for computer vision processing methods powered by an image recognition tool before concolic testing via a data-driven approach to the code obtained through AST-based code transformation.
The target functions are concolic tested independently on unnecessary parameters to the designated variables.
However, for a practical application such as reactive systems where a target function might be called repeatedly, the concolic execution should retain the values or results of the previous processes and pass them down to the next execution.
The state extractor can address its problem by assigning all expected patterns of state variables and their values to the target function.
Concolic testing runs on each combination of the patterns of state variables and the number of images in the video used in the tests, thereby sampling the state transition paths.

\subsection{Model Generator}
In the Model Generator, a state set $S$ and a transition set $R$ of the Kripke structure are generated from the strings aggregated during concolic testing using regular expressions.
\color{black}
\begin{algorithm}
    \caption{Model construction from symbolic execution sequences}
    \label{alg1}
\SetAlgoLined
\KwData{symbolic execution sequences \textit{QUEUE}}
\KwData{a set of states $S$}
\KwData{a set of transitions $R$}
\KwData{a set of labels $L$}
\KwInit{\\
\vspace{0.5mm}
$\begin{array}{@{}c@{\ }c@{\ }c}
    S & \leftarrow & \emptyset;\\
    R & \leftarrow & \emptyset;\\
    L & \leftarrow & \emptyset;\\
\end{array}$
}
\While{\textit{QUEUE} is not empty}{
    \textit{sequence} $\leftarrow$ \textit{QUEUE}.pop()\;
    \ForEach{$r \in sequence$}{
        $s, s\prime \leftarrow r$\;
        \If{$s \notin S$ \textbf{or} $s\prime \notin S$}{
            $S \leftarrow S \cup \{s, s\prime\}$\;
            $L \leftarrow L \cup \{\phi(s), \phi(s\prime)\}$\;
        }
        \If{$r \notin R$}{
            $R \leftarrow R \cup \{r\}$\;
        }
    }
}
\end{algorithm}
Algorithm \ref{alg1} shows how to construct a Kripke structure from symbolic execution sequences generated by PyExZ3.
Let $S$ be a set of states, $R \subseteq S\times S$ be a set of relations, $L$ be a set of labels.
Let $\phi$ be a labeling function that maps each state $s$ to a label.

By using a thread-safe data structure for \textit{QUEUE} in Algorithm \ref{alg1}, the state extractor and the constructor can be executed in a multi-process environment.

%% file: Ch4_Impl_Details_HO.tex
\section{Implementation Details}
\label{sec:4}
\subsection{State Extractor}
As the first step of CodoMo system, the state extractor mainly plays the following roles:
\begin{itemize}
\setlength{\itemsep}{1mm}
\item extracting functions or methods which are to be tested;
\item analyzing the abstract syntax tree (AST) of them to detect and highlight Assign nodes or If nodes concerned with designated variables and attributes;
\item loading external test data for symbolic execution of extracted functions if needed;
\item making minor modification required for executing PyExZ3.
\end{itemize}

The state extractor in CoDoMo introduces two key features for utilizing PyExZ3. The first is its ability to handle object-oriented programs.
PyExZ3 can only accept functions in a given code as its targets and thus, any class methods cannot directly undergo concolic testing.
The state extractor address this problem by transforming designated methods into suitable functions without altering context.
One can see from Figs.~\ref{fig:snipped_code1}, \ref{fig:snipped_code2} class attributes \texttt{self.current\_state} are replaced with variables \texttt{current\_state\_\_Tello\_TEST} where ``Tello\_TEST'' is a name of an object class that these attributes belong to.

The second feature is to explore all possible substates along every path.
After we designate some class attributes, the state extractor detects assignments and conditional branches related to the attributes.
In this process, they are marked by the state extractor inserting print statements: wrapping the body of detected branches with ``\texttt{[BEGIN IF]}'' and ``\texttt{[END IF]}''; dumping the values of the attribute before and after the detected assignments with a right arrow ``\texttt{->}''
%In this way, the state extractor implemented in CodoMo detects all assignments of values or the branches involving the designated attributes inserting print statements.
(compare Fig.~\ref{fig:snipped_code1} with Fig.~\ref{fig:snipped_code2}). 
These inserted statements are analyzed via regular expressions to record state transitions together with the results of the concolic testing by PyExZ3.
\par

\begin{figure}[ht]
    \centering
    \begin{ttfamily}
    {\fontsize{8pt}{4mm}\selectfont
    \begin{tabular}{|@{\ }c@{\ }p{73mm}@{\ }|}\hline
        274 & if self.current\_state != (TelloState.LAND or TelloState.FIN) :\\
        275 & ~~~~self.current\_state = TelloState.OPEN\\
        \hline
    \end{tabular}
    }
    \end{ttfamily}
    \caption{A snippet of an original code}
    \label{fig:snipped_code1}\par
    \vspace{10mm}
    \begin{ttfamily}
    {\fontsize{8pt}{4mm}\selectfont
    \begin{tabular}{|@{\ }c@{\ }p{73mm}@{\ }|}\hline
        225 & if current\_state\_\_Tello\_TEST != (TelloState.LAND or TelloState.FIN) :\\
        226 & ~~~~\textbf{print('[BEGIN IF]')}\\
        227 & ~~~~print(current\_state\_\_Tello\_TEST)\\
        228 & ~~~~\textbf{print('->')}\\
        229 & ~~~~current\_state\_\_Tello\_TEST = TelloState.OPEN\\
        230 & ~~~~print(current\_state\_\_Tello\_TEST)\\
        231 & ~~~~\textbf{print('[END IF]')}\\
        \hline
    \end{tabular}
    }
    \end{ttfamily}
    \caption{A snippet of the transformed code}
    \label{fig:snipped_code2}
\end{figure}

\subsection{Multiprocessing of State Extractor and Model Generator}

On-the-fly execution can tackle the state explosion problem by searching a state space and generating states in parallel.
In model checking, on-the-fly execution refers to an algorithm where the state space is generated dynamically.
We envision accelerating model checking in CodoMo tool through on-the-fly execution.

\begin{figure}[h]
\begin{center}
  \includegraphics[width=\linewidth]{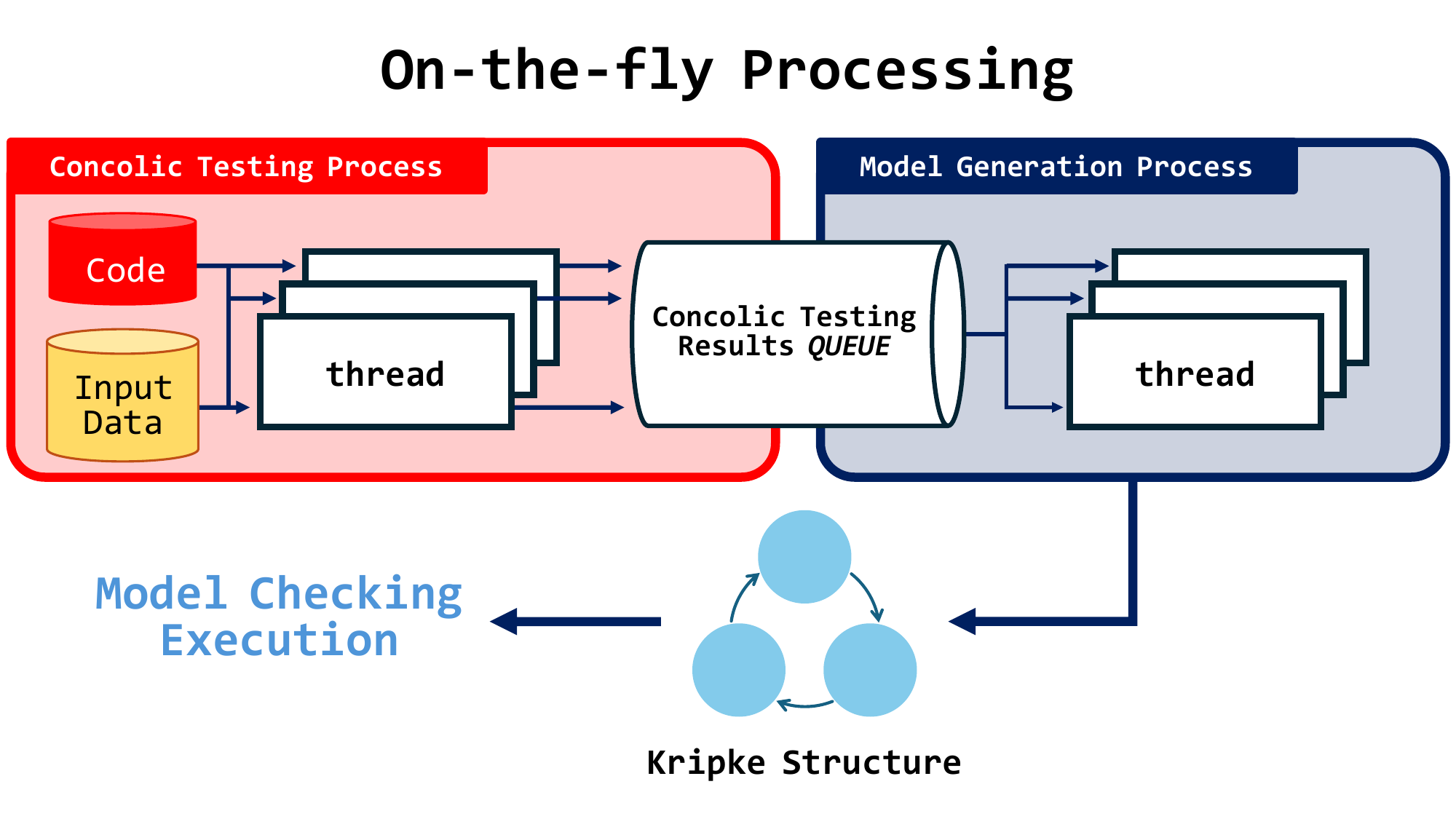}
  \caption{On-the-fly multiprocessing architecture of the concolic testing process and the model generation process}
  \label{figure2}
  \end{center}
\end{figure}

As a result of sequential experiments of execution, it was found that the execution time of the concolic testing process got significantly longer compared to that of the model generation process.
Therefore, PyExZ3 should be working on the concolic testing independently while the model generator and pyModelChecking are temporary idle.
We adopt a multiprocessing approach that integrates the model generation process with an on-the-fly model checker, as shown in Fig.~\ref{figure2}.
To be honest, the benefits of this approach cannot be granted immediately, since pyModelChecking executed after these processes (see Fig.~\ref{figure2}) does not support on-the-fly execution.
We are considering the adoption of alternative model checkers, particularly from a perspective of effectively showing counterexamples in model checking.

%% file: Ch5_Result_Discussion_HOG.tex
\section{Results and Discussions}
\label{sec:5}

\begin{table*}[!h] % ここで table* 環境を使用
\newcommand{\bhline}[1]{\noalign{\hrule height #1}} 
\centering
\caption{CTL* FORMULA REPRESENTATION}
\renewcommand{\arraystretch}{1.3}
{\fontsize{7pt}{3.5mm}\selectfont%\scriptsize
\begin{tabular}{ccl@{\ }c}
\bhline{0.8pt}
Symbol & Property & \multicolumn{1}{c}{Formula} & \begin{tabular}{c}Verification\\Result\end{tabular} \\
\bhline{0.8pt}
$\phi_1$ & Halt property & $\mathbf{A}\mathbf{G}\left(\left(\texttt{current\_state} = \textit{TAKEOFF}\right) \rightarrow \mathbf{F} \left(\texttt{current\_state} = \textit{LAND}\right)\right)$ & False \\
$\phi_2$ & Flip Action property (R) &
$\mathbf{E}\left(\mathbf{F}\left(\neg\left(\texttt{current\_state} = \textit{FIN}\right) \rightarrow \left( \mathbf{X}\left(\texttt{current\_state} =\textit{OPEN}\right) \land \texttt{current\_state} =\textit{RIGHT}\right)\right)\right)$
& True \\
$\phi_3$ & Flip Action property (L)  &
$\mathbf{E}\left(\mathbf{F}\left(\neg\left(\texttt{current\_state} = \textit{FIN}\right) \rightarrow \left( \mathbf{X}\left(\texttt{current\_state} =\textit{OPEN}\right) \land \texttt{current\_state} =\textit{LEFT}\right)\right)\right)$
& True \\
$\phi_4$ & Weak Reachability  &
\scalebox{1}{$\neg ( \texttt{current\_state} =\textit{FIN}) \rightarrow \mathbf{E}( \mathbf{F}( \texttt{current\_state} =\textit{FIN} ))$}
& True \\
$\phi_5$ & Strong Reachability  &
$\neg \left(\texttt{current\_state} =\textit{FIN}\right) \rightarrow \mathbf{A}\left(\mathbf{G}\left(\mathbf{F}\left( \texttt{current\_state} =\textit{FIN} \right)\right)\right)$
& False \\
$\phi_6$ &Irreversibility &
$\mathbf{A}\mathbf{G}\left(\left(\texttt{current\_state} = \textit{TAKEOFF}\right)\rightarrow \neg \left( \texttt{current\_state} = \textit{LAND}\right)\right)$
& True\\
\bhline{0.8pt}
\end{tabular}
}
\renewcommand{\arraystretch}{1}
\label{t1}
\end{table*}

\subsection{Verification Example: Drone hand gesture operation}

CodoMo was tested on a toy drone programming for hand gesture operation.
It is designed for guest lectures aiming at raising awareness of AI and IoT technologies among high school students.
We visited 16 high schools to deliver lectures and group works by using the program for more than 300 students many of whom were unfamiliar with either drone controlling or computer programming. 
Our drones are programmed to accept six gestures (up, land, left, right, open, gal) by right fingers with cameras attached to the front side and select one of eight states (\textit{TAKEOFF}, \textit{LAND}, \textit{FIN}, \textit{LEFT}, \textit{RIGHT}, \textit{UP}, \textit{OPEN}, \textit{GAL}).\par
Drones start up with the \textit{LAND} state. They \textit{TAKEOFF} by ``up'' gestures. Drones in the air can move \textit{RIGHT}, \textit{LEFT}, or \textit{UP} by ``right'', ``left'', or ``up'' gestures, respectively.
``Open'' gesture makes drones \textit{OPEN}, that is, ready for flipping and waiting ``right'' or ``left'' gestures. \textit{GAL} is an independent joke state following from ``gal'' gestures.
Finally, drones may be grounded by ``land'' gestures and end up with \textit{FIN}.\par

From a safety perspective, toy drones must be under control during programming conducted by student, even if gestures are incorrectly interpreted.
CodoMo verifies the control program to secure learning environment for students.
Precisely speaking, CodoMo checks the following properties of the system.
\par

\subsubsection{Halt Property}
The drone system must transit to the stopped state (\textit{FIN}) after flying.
We set the halt property described the CTL* specification $\phi_1$ in Table \ref{t1}.
Here, \texttt{current\_state} is a state variable defined as an initial member variable in the Python class.

\subsubsection{Flip Action Property}
This gesture-based drone interfacing system is designed to enable easy operation using simple hand gestures without adding complexity.
For example, the gesture that commands the drone to perform a ``flip'' action, which involves a 360-degree rotation in any direction, is implemented by combining existing gestures: pointing in the desired direction after opening the hand.
We defined the flip actions for each direction as flip action properties $\phi_2$ and $\phi_3$ for left and right flips, respectively, and verify whether there are paths that execute these flip actions.
\subsubsection{Reachability}
Reachability is a property that indicates whether a specific state can be reached. In the drone system, it is desirable to reach the \textit{FIN} state eventually.
We defined weak reachability, $\phi_4$, which shows that at least one path exists that leads to the \textit{FIN} state, and strong reachability, $\phi_5$, which ensures that every path inevitably leads to the \textit{FIN} state.
\subsubsection{Irreversibility}
Irreversibility is also crucial in the context of temporal constraints, as it ensures that once a process advances beyond a certain point, it cannot revert, thereby maintaining the integrity of time-dependent operations.
By explicitly defining distinct initial and terminal states in a reactive system, looping transitions can be eliminated, establishing a solid foundation for verifying boundedness and fairness.
We defined the Irreversibility property $\phi_6$, which states that once the system reaches \textit{TAKEOFF}, it will never return to \textit{LAND}.

\subsection{Experiments}

\begin{figure}[h] % 図1
    \centering
    \includegraphics[width=\linewidth]{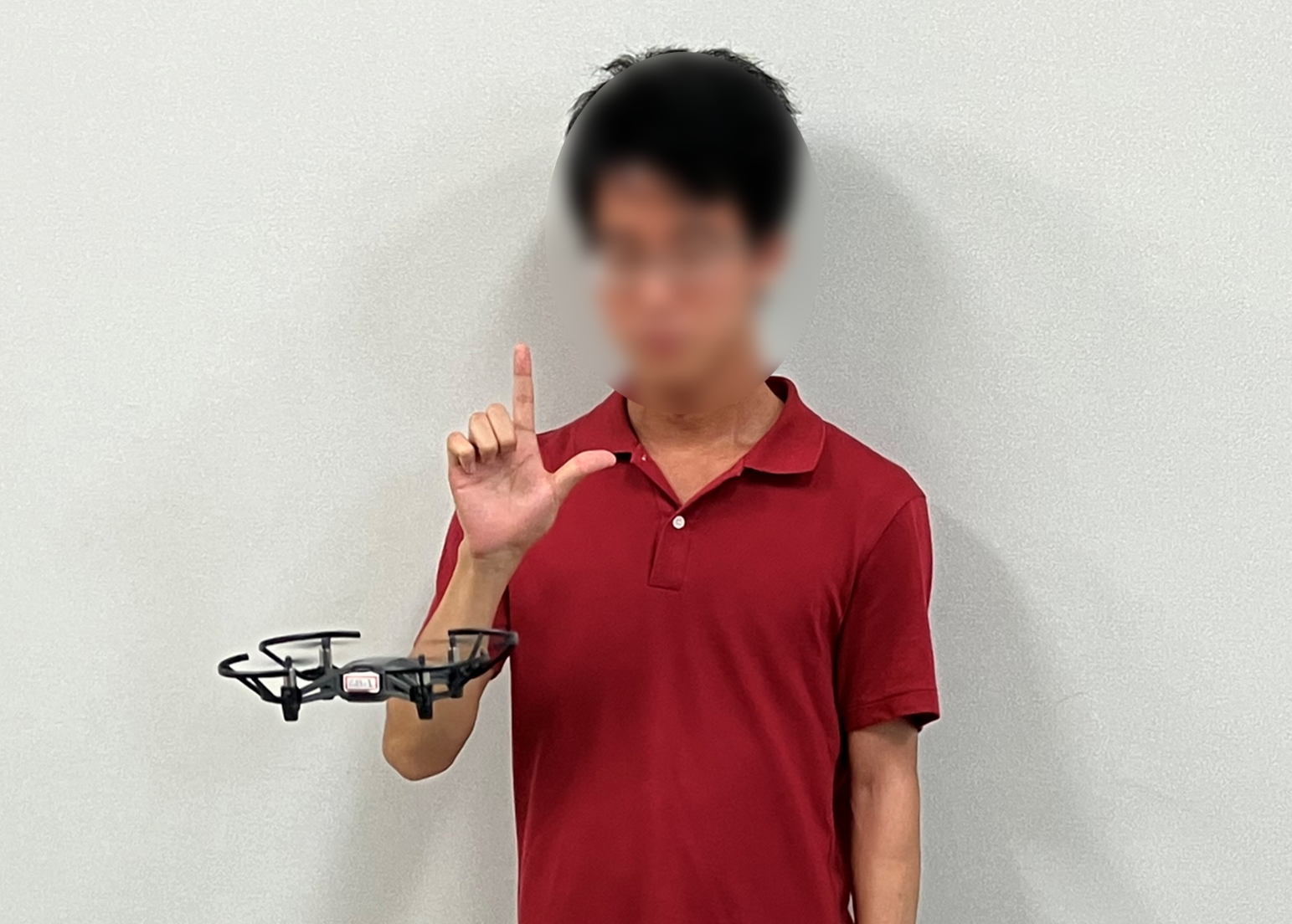}
    \caption{Experimental setup for drone gesture control}
    \label{fig:ex1}
\end{figure}

\begin{figure}[h] % 図2
    \centering
    \includegraphics[width=\linewidth]{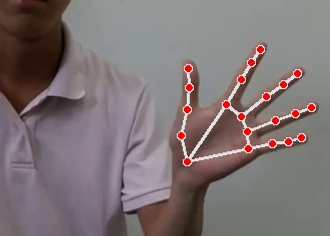}
    \caption{Skeleton extraction process for hand gesture recognition}
    \label{fig:ex2}
    \vspace{-0.5cm}
\end{figure}

Figs.~\ref{fig:ex1}--\ref{fig:ex2} show scenes of experiments where two subjects performed gesture-based drone operations, with the drone capturing approximately one minute of video footage each, resulting in 224 sample images.
Concolic testing was conducted using PyExZ3 on drone interfacing program Tello\_finger\_sign.py made by us \cite{harie2023} together with these 224 images, specifically using \texttt{mp\_camera} method of \texttt{Tello\_TEST} class as the target function.

Fig.~\ref{fig:appendix} shows a part of the intermediate files created during state space generation for model checking, produced by the state extractor.
{\sloppy%行間詰め
In Fig.~\ref{fig:appendix}, it can be observed that through the concolic testing by PyExZ3, \texttt{index\_finger\_left\_frams\_\_Tello\_TEST=10} is solved (line 1239), leading to the execution of a path where a state transition occurs from \textit{TAKEOFF} to \textit{LEFT} (obviously confirmed by lines 1246--1248).
Furthermore, \texttt{index\_finger\_right\_frams\_\_Tello\_TEST=14} is solved (line 1251), leading to the execution path where a state transition occurs from \textit{TAKEOFF} to \textit{RIGHT} (see lines 1258--1260).%
}%

\begin{figure}[h]
\centering
  \includegraphics[width=\linewidth]{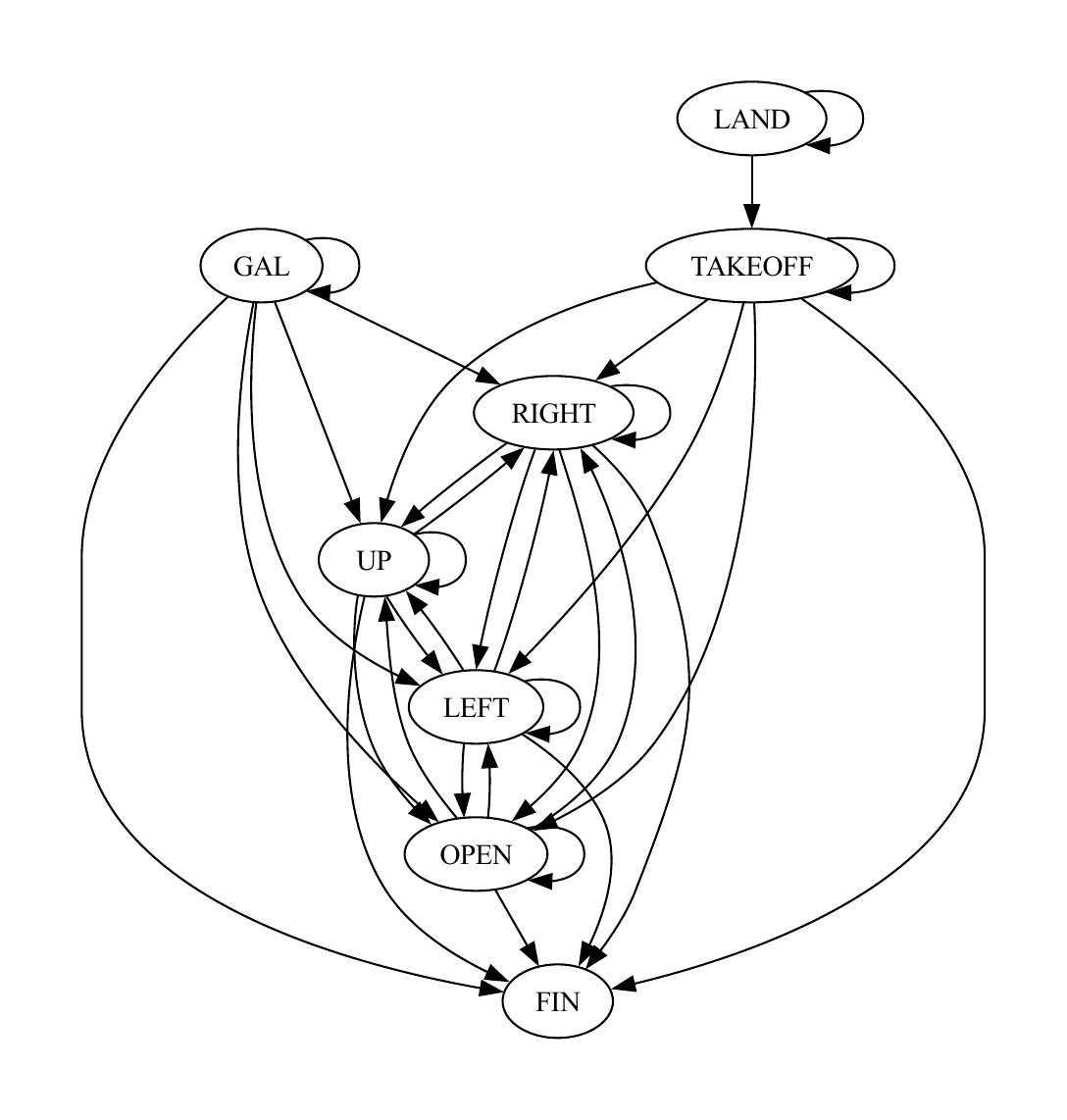}
  \caption{The state space represented by the Kripke structure obtained through MDRE from the Python code and image datasets.}
  \label{fig:ss}
\end{figure}

This process ultimately resulted in the state space with 8 states and 34 transitions shown in Fig.~\ref{fig:ss}.

Model checking was then performed on the state space based on the verification properties defined in the previous section, and the results are presented in TABLE~\ref{t1}.

% \begin{table*}[!h]
% \centering
% \begin{minipage}{\textwidth}
% \newcommand{\bhline}[1]{\noalign{\hrule height #1}} 
% \centering
% \caption{Code Differences}
% \begin{tabular*}{\textwidth}{@{\extracolsep{\fill}}@{\hspace{5mm}}ccccc@{\hspace{5mm}}}\bhline{0.8pt}
% Comparison & Difference in lines & Difference in bytes & Change rate & Levenshtein distance\\\bhline{0.8pt}
% OC with PTC & 532 lines     & 33206 bytes   & 163.19\%    & 5783 characters\\
% PTC with FTC* & 20 lines      & 1276 bytes    & 6.86\%      & 1364 characters\\
% PTC with FTC & 40 lines      & 2401 bytes    & 13.70\%     & 1373 characters\\
% \bhline{0.8pt}
% \end{tabular*}
% \label{table:code_diffs}
% \end{minipage}
% \end{table*}

\vspace{0.3cm} % Space between tables
\begin{figure*}[!h]
    \centering
    {\fontsize{8pt}{4mm}\selectfont
    \begin{ttfamily}
    \begin{tabular}{|cp{160mm}|}\hline
    1233 & Exploring dummy\_code.mp\_camera\_\_Tello\_TEST\\
    1234 & [('current\_state\_\_Tello\_TEST', 0), ('index\_finger\_right\_frames\_\_Tello\_TEST', 0), ('index\_finger\_left\_frames\_\_Tello\_TEST', 0), ('index\_finger\_open\_frames\_\_Tello\_TEST', 0), ('gal\_finger\_frames\_\_Tello\_TEST', 0), ('index\_finger\_down\_frames\_\_Tello\_TEST', 0), ('index\_finger\_up\_frames\_\_Tello\_TEST', 0)]\\
    1235 & [initialize]\\
    1236 & TelloState.TAKEOFF\\
    1237 & [initialize]\\
    1238 & TelloState.TAKEOFF\\
    \textbf{1239} & [('current\_state\_\_Tello\_TEST', 0), ('index\_finger\_right\_frames\_\_Tello\_TEST', 0), \textbf{('index\_finger\_left\_frames\_\_Tello\_TEST', 10)}, ('index\_finger\_open\_frames\_\_Tello\_TEST', 0), ('gal\_finger\_frames\_\_Tello\_TEST', 0), ('index\_finger\_down\_frames\_\_Tello\_TEST', 0), ('index\_finger\_up\_frames\_\_Tello\_TEST', 0)]\\
    1240 & [initialize]\\
    1241 & TelloState.TAKEOFF\\
    1242 & [initialize]\\
    1243 & [BEGIN IF]\\
    1244 & [END IF]\\
    1245 & [BEGIN IF]\\
    1246 & TelloState.TAKEOFF\\
    1247 & ->\\
    1248 & TelloState.LEFT\\
    1249 & [END IF]\\
    1250 & TelloState.LEFT\\
    \textbf{1251} & [('current\_state\_\_Tello\_TEST', 0), \textbf{('index\_finger\_right\_frames\_\_Tello\_TEST', 14)}, ('index\_finger\_left\_frames\_\_Tello\_TEST', 0), ('index\_finger\_open\_frames\_\_Tello\_TEST', 0), ('gal\_finger\_frames\_\_Tello\_TEST', 0), ('index\_finger\_down\_frames\_\_Tello\_TEST', 0), ('index\_finger\_up\_frames\_\_Tello\_TEST', 0)]\\
    1252 & [initialize]\\
    1253 & TelloState.TAKEOFF\\
    1254 & [initialize]\\
    1255 & [BEGIN IF]\\
    1256 & [END IF]\\
    1257 & [BEGIN IF]\\
    1258 & TelloState.TAKEOFF\\
    1259 & ->\\
    1260 & TelloState.RIGHT\\
    1261 & [END IF]\\
    1262 & TelloState.RIGHT\\
    \hline
    \end{tabular}
    \end{ttfamily}
    }
    \caption{Results of the concolic execution by PyExZ3}
    \label{fig:appendix}
\end{figure*}

\subsection{Discussions}

\begin{table*}[!h]
\centering
\begin{minipage}{\textwidth}
\newcommand{\bhline}[1]{\noalign{\hrule height #1}} 
\centering
\caption{Code Differences}
\begin{tabular*}{\textwidth}{@{\extracolsep{\fill}}@{\hspace{5mm}}ccccc@{\hspace{5mm}}}\bhline{0.8pt}
Comparison & Difference in lines & Difference in bytes & Change rate & Levenshtein distance\\\bhline{0.8pt}
OC with PTC & 532 lines     & 33206 bytes   & 163.19\%    & 5783 characters\\
PTC with FTC* & 20 lines      & 1276 bytes    & 6.86\%      & 1364 characters\\
PTC with FTC & 40 lines      & 2401 bytes    & 13.70\%     & 1373 characters\\
\bhline{0.8pt}
\end{tabular*}
\label{table:code_diffs}
\end{minipage}
\end{table*}

As an overview of the CodoMo process, TABLE~\ref{table:code_diffs} compares the original code (OC) a primal transformed code (PTC), and a finally transformed code (FTC).
PTC is generated automatically through AST analysis in the first step of the state extractor. The difference from OC in TABLE~\ref{table:code_diffs} shows PTC contains a lot of the print statements inserted as markers for tracking designated attributes thereby clearly marking the state transitions. FTC is derived from PTC through manual processes roughly classified into two factors: deactivating specific function calls that affect PyExZ3 execution; making intentional modifications based on the context or necessary corrections of inappropriate structures.
The first one is possibly automated in a future work, whereas the last one is considered as a ``critical'' factor.
FTC* in TABLE~\ref{table:code_diffs} includes PTC itself modified according to the first factor.
Therefore, by the second and third rows, the critical factor occupies 20 lines, 1125 bytes, 6.84\%, 9 characters, which is for the most part commenting some rows out.

The model checking violations for the generated state space include the Halt property $\phi_1$ and Strong Reachability $\phi_5$. 
These counterexamples are straightforward and include trivial cases such as self-loops in states other than the \textit{FIN} state.
This is because the drone program under verification is a reactive system, designed to accept an infinite sequence of inputs (gestures), which is appropriate for its intended function.
However, given constraints such as battery capacity and the time limitations for educational purposes, it is also reasonable to impose conditions that result in a bounded execution trace.

By managing not only the drone's flight states but also the battery capacity (which can be updated or retrieved through communication with the Tello drone program) and the number of state transitions triggered by gesture operations, it is possible to improve the system to meet the aboved specifications.
However, such improvements, which increase the combination of internal states, could lead to state explosion.
There have been reports that pyModelChecking, surprisingly, outperforms LTSmin in efficiency for small examples.
pyModelChecking performs well with smaller inputs, despite their algorithms not operating on-the-fly; however, they struggle with larger problems, failing to terminate within reasonable time and memory constraints.
If the need for on-the-fly execution arises, it may be necessary to introduce an alternative model checker that can run on Python, other than pyModelChecking.

%% file: Ch6_Conclusion_HOG.tex
\section{Conclusion}
\label{sec:6}

In this study, we have successfully semi-automated the process of extracting models from the system implementation and conducting verification on generated models. 
Using the CodoMo system architecture, we enabled flexible verification of legacy systems not initially developed with MDD approach.
Following the verification of a drone system using computer vision technology, we empirically demonstrated the potential of a data-driven approach for state space generation and the applicability of formal methods to machine learning systems.
However, a fundamental challenge in this study is the lack of exhaustiveness in the input images provided for verification. 
Moving forward, we aim to apply this flexible approach of model generation and verification from systems to various machine learning models developed in our previous works \cite{harie2021, harie2023_a}, and further establish CodoMo system.
In the future, incorporating internal variables like battery status may require integrating another Python-based model checker into CodoMo.

%% file: Ch7_Acknowledgement.tex
\section{Acknowledgments}
We would like to express our special thanks of gratitude to Ryosuke Ono for his invaluable assistance in conducting the experiments. %
This work was supported by Japan Society for the Promotion of Science (JSPS) KAKENHI Grant Number 23K05416 (Grant-in-Aid for Scientific Research(C)).